\documentclass[11pt,a4paper]{article}
\usepackage[latin1]{inputenc}
\usepackage[english]{babel}

\usepackage{amsfonts}
\usepackage{amsmath}
\usepackage{amssymb}
\usepackage{amsthm}
\usepackage{bm}
\usepackage[round]{natbib}
\usepackage{graphicx}
\usepackage{caption}
\usepackage{subcaption}
\usepackage{float}
\usepackage[parfill]{parskip}
\usepackage{epstopdf}
\usepackage{appendix}
\usepackage{booktabs}
\usepackage{verbatim}
\usepackage{url}

\newcommand{\dd}{{\mathrm d}}

\newcommand{\e}{{\rm e}}
\newcommand{\E}{{\mathbb E}}

\newcommand{\Q}{{\mathbb Q}}

\newcommand{\R}{{\mathbb R}}

\newcommand{\Fcal}{{\mathcal F}}

\newcommand{\Ocal}{{\mathcal O}}

\makeatletter
\def\thm@space@setup{%
  \thm@preskip=\parskip \thm@postskip=0pt
}
\makeatother

\newtheorem{proposition}{Proposition}[section]

\newtheorem{theorem}[proposition]{Theorem}

\newtheorem{remark}[proposition]{Remark}

\title{SABR smiles for RFR caplets}
 \author{Sander Willems\\ NatWest Markets\footnote{The statements and opinions expressed in this article are my own and do not represent the views of NatWest Markets Plc, NatWest Markets N.V.\ (and/or any branches) and/or their affiliates.}}
\date{First version: April 3, 2020\\ This version: May 05, 2020}

\begin{document}

\maketitle

\begin{abstract}
We present a natural extension of the SABR model to price both backward and forward-looking RFR caplets in a post-Libor world. Forward-looking RFR caplets can be priced using the market standard approximations of \cite{hagan2002managing}. We provide closed-form effective SABR parameters for pricing backward-looking RFR caplets. These results are useful for smile interpolation and for analyzing backward and forward-looking smiles in normalized units.\end{abstract}

\section{Introduction}
Following a speech by the Financial Conduct Authority in July 2017, it became apparent that Libor is expected to cease after end 2021. Regulators have actively pushed for the development of new interest rate benchmarks that are firmly anchored to actual transactions, rather than expert judgement of a handful of panel banks. These new benchmarks are generally referred to as RFRs, short for risk-free rates. We refer to them in this paper simply as overnight rates, since that is what they are based on. Libor is currently by far the most important interest rate benchmark. The transition away from Libor therefore has widespread consequences for financial products, see for example \cite{henrard2019libor} for a quantitative finance perspective. We focus in this paper in particular on the consequences for interest rate caps and floors. 

A standard Libor caplet pays the difference between a Libor rate and a strike rate with a floor at zero. The payment occurs at the end of the accrual period to which the Libor rate refers, while the Libor rate itself, and therefore the caplet's payoff, is already known at the start of the accrual period. In a post-Libor world, caplets will possibly exist in two different formats. The first one, which we refer to as the \textit{backward-looking} caplet, replaces the Libor rate by the compounded overnight rate over the accrual period. This product is conceptually different from a Libor caplet, since its payoff is only known at the end of the accrual period. The second one, which we refer to as the \textit{forward-looking} caplet, replaces the Libor rate by the par rate of a single-period Overnight Index Swap (OIS) over the accrual period. This one is conceptually similar to a Libor caplet, but it requires a reliable OIS rate benchmark to be developed.\footnote{In January 2020, the ICE Benchmark Administration launched a market consultation on the introduction of an ICE swap rate based on SONIA, see \cite{ice2020sonia}. }

A natural modelling approach when working with overnight rates is to specify dynamics for the short-rate. \cite{henrard2004overnight,henrard2006skewed,henrard2019quant} and \cite{turfus2020caplet} derive explicit formulas for pricing backward-looking caplets in a single factor Gaussian short-rate model. These type of models however fail to produce a smile and can therefore only be calibrated to at-the-money caplets. This limits their practical use for traders. \cite{macrina2020rational} work with a linear-rational framework, similar to the framework in \cite{filipovic2019term}, and rely on affine transforms to obtain semi-closed form solutions for backward-looking caplet prices. In a seminal contribution, \cite{lyashenko2019looking} extend the Libor Market Model (LMM) for use with compounded overnight rates, which they refer to as the Forward Market Model (FMM). They provide explicit formulas for backward-looking caplets for the lognormal case with constant volatility, which again does not produce a smile. The FMM framework is however much more general and allows in particular for stochastic volatility dynamics as well, although calibration can prove difficult if analytic solutions for backward-looking caplet prices are not available.

In this paper we propose an extension of the SABR model introduced by \cite{hagan2002managing}. Similarly as in \cite{lyashenko2019looking}, we specify dynamics for the forward value of the compounded overnight rate over a certain accrual period. Before entering the accrual period, this rate follows standard SABR dynamics. Forward-looking caplets, whose payoff is known at the start of the accrual period, can therefore be priced using the market standard implied volatility approximation of \cite{hagan2002managing}. Once the rate enters the accrual period, it keeps evolving stochastically but its volatility is gradually scaled down to account for already realized overnight fixings. We introduce a new parameter that controls how fast the volatility is reduced and show that it can be related to the mean-reversion speed in a short-rate model. We build on the results of \cite{hagan2018managing} to derive effective constant SABR parameters that account for this time-dependent volatility scaling in the accrual period. Backward-looking caplets, which expire at the end of the accrual period, can therefore again be priced using the implied volatility approximation of \cite{hagan2002managing}, but with different parameters than forward-looking caplets. Our results allow traders to think of backward and forward-looking smiles in the same units, since we can map forward-looking SABR parameters into backward-looking ones and vice versa. \cite{piterbarg2020interest} highlights the importance of such an exercise and derives himself a correction to the initial volatility parameter in the SABR model, leaving the smile parameters unchanged. We show that this is indeed an important parameter to make an adjustment for in order to price the at-the-money backward-looking caplet correctly. For strikes away from the money, we show that an adjustment to the volatility-of-volatility parameter is important as well, in particular if we are close to or already in the accrual period. 

The remaining of this paper is structured as follows. Section \ref{sec:rfr_caplets} describes the payoff of backward and forward-looking caplets. Section \ref{sec:dynamics} introduces the backward-looking SABR model. Section \ref{sec:effective} derives effective SABR parameters, which is the main contribution of this paper. Section \ref{sec:numerical} contains a numerical study of the new model. Section \ref{sec:conclusion} concludes. All proofs and auxiliary results are collected in the Appendix.

\section{Caplets in a post-Libor world}\label{sec:rfr_caplets}
Consider two times $\tau_0<\tau_1$ and define the continuously\footnote{In practice, overnight rates are compounded discretely on a daily basis, rather than continuously. For the purpose of this paper, however, this distinction is only of minor importance.} compounded overnight rate as
\begin{align*}
    R = \frac{1}{\tau_1-\tau_0}\left(e^{\int_{\tau_0}^{\tau_1} r(s)\,\dd s}-1\right),
\end{align*}
where $r(t)$ denotes an instantaneous proxy of the overnight rate. We define for $t\le \tau_1$
\begin{align}
    R(t)= \E^{\tau_1}_t[R],
    \label{eq:def_Rc}
\end{align}
where $\E^{\tau_1}_t[\cdot]$ denotes the $\Fcal_t$-conditional expectation with respect to the $\tau_1$-forward measure $\Q^{\tau_1}$. In other words, a swap exchanging a floating payment $R$ at $\tau_1$ against a fixed payment $K$ will have zero value at time $t$ if $K=R(t)$. Note that $R(\tau_1)=R$, since the random variable $R$ is $\Fcal_{\tau_1}$-measurable.

In a post-Libor world, there will possibly be two types of caplets. The first type, which we refer to as the backward-looking caplet, pays at time $\tau_1$
\begin{align*}
    V^b_{cpl}(\tau_1) = (R(\tau_1)-K)^+,
\end{align*}
where $K$ is the strike rate. This payoff is only known at payment time $\tau_1$. The second type, which we refer to as the forward-looking caplet, pays at time $\tau_1$
\begin{align*}
    V^f_{cpl}(\tau_1) = (R(\tau_0)-K)^+.
\end{align*}
The payoff of the forward-looking caplet is already known at time $\tau_0$, but only paid at time $\tau_1$. The forward-looking caplet is therefore very similar to a traditional Libor caplet and equivalent to a single-period OIS swaption (see Remark \ref{remark:swpt}). The value at time $t\le \tau_1$ of the backward and forward-looking caplets are respectively given by:
\begin{align*}
    V_{cpl}^b(t) &= P(t,\tau_1)\E^{\tau_1}_t[(R(\tau_1)-K)^+],\\
     V_{cpl}^f(t) &= P(t,\tau_1)\E^{\tau_1}_t[(R(\tau_0)-K)^+],
\end{align*}
where $P(t,\tau_1)$ denotes the price at time $t$ of a zero-coupon bond with maturity $\tau_1$.

From Jensen's inequality, we get for $t\le \tau_0$
\begin{align}
    V_{cpl}^f(t) 
    &\le P(t,\tau_1)\E^{\tau_1}_t\left[\E_{\tau_0}^{\tau_1}\left[(R(\tau_1)-K)^+\right] \right]=  V_{cpl}^b(t).
    \label{eq:jensen}
\end{align}
Therefore, the backward-looking caplet is always worth at least as much as the forward-looking caplet, provided we have not yet entered the accrual period. The above inequality has been highlighted by several authors, e.g., \cite{lyashenko2019looking}, \cite{willems2019}, and \cite{piterbarg2020interest}. Note that we have equality in \eqref{eq:jensen} if and only if interest rates are deterministic in the accrual period.

\begin{remark}\label{remark:swpt}
By changing to the $\tau_0$-forward measure, we can express the value at time $t\le \tau_0$ of the forward-looking caplet as
\begin{align*}
    V_{cpl}^f(t) &= P(t,\tau_1)\E^{\tau_1}_t[(R(\tau_0)-K)^+]\\
    & = P(t,\tau_0)\E^{\tau_0}_t[P(\tau_0,\tau_1)(R(\tau_0)-K)^+].
\end{align*}
Note that $P(\tau_0,\tau_1)(R(\tau_0)-K)$ corresponds to the value at time $\tau_0$ of a spot starting single-period OIS exchanging at time $\tau_1$ a floating payment $R=R(\tau_1)$ against a fixed payment $K$. Therefore, the forward-looking caplet and the single-period OIS swaption have identical present values at time $t\le \tau_0$. Note that this is not true for the backward-looking caplet.
\end{remark}

\section{Backward-looking SABR model} \label{sec:dynamics}
From its definition in \eqref{eq:def_Rc}, we require $R(t)$ to be a $\Q^{\tau_1}$-martingale. Within the accrual period, i.e., for $t\in[\tau_0,\tau_1]$, $R(t)$ contains an increasing number of realized rates and this should be reflected in its volatility. With these considerations in mind, we propose the following dynamics for $R(t)$:
\begin{align}
    &\dd R(t) = \psi(t)\sigma(t) R(t)^\beta  \,\dd B(t), \quad\psi(t)=\min\left(1,\frac{\tau_1 - t}{\tau_1 - \tau_0}\right)^q , \label{eq:spec_R}\\
    &\dd \sigma(t) = \nu \sigma(t)\, \dd W(t),\quad \sigma(0)=\alpha, \label{eq:spec_sig}
\end{align}
with $q>0$, $\beta\in [0,1]$, $\alpha,\nu>0$, and $B(t),W(t)$ two $\Q^{\tau_1}$-Brownian motions with $\dd B(t)\, \dd W(t) = \rho\, \dd t$, $\rho\in[-1,1]$.\footnote{We assume a zero lower bound for $R(t)$, however it is straightforward to accommodate an arbitrary lower bound through a displacement.} For $t\le \tau_0$, $R(t)$ follows standard SABR dynamics. For $t\ge \tau_0$, the stochastic volatility process is scaled down so that $R(t)$ becomes increasingly deterministic for $t\to\tau_1$. The newly introduced parameter $q$ controls how fast the volatility is reduced. Larger values lead to a faster reduction, see Figure \ref{fig:psi_q} for an illustration. We show in Appendix \ref{app:short-rate} that $q-1$ can be related to the mean-reversion speed of the overnight rate. 

\begin{figure}
    \centering
    \includegraphics[width = 0.8\textwidth]{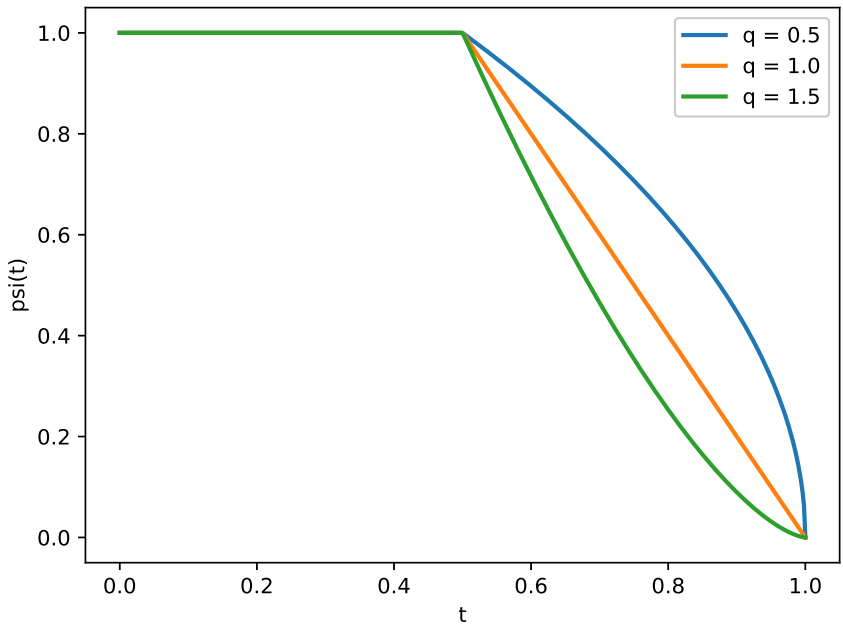}
    \caption{Function $\psi(t)$ for different values of $q$ with $\tau_0=0.5$ and $\tau_1 = 1$.}
    \label{fig:psi_q}
\end{figure}

Pricing forward-looking caplets, i.e., options on $R(\tau_0)$, can be done easily using the implied volatility approximations of \cite{hagan2002managing}:
\begin{align}
    &V_{cpl}^f(0) = P(0,\tau_1) \pi\left(\tau_0,K,  R(0), \sigma^f_{IV}\right),\label{eq:black_fwd}\\
    &\sigma^f_{IV} \approx \sigma_{hagan}(\tau_0,K,R(0),\alpha,\beta,\rho,\nu),\label{eq:hagan_fwd}
\end{align}
where $\pi(\cdots)$ denotes \cite{black1976pricing}'s formula  and $\sigma_{hagan}(\cdots)$ denotes the Black implied volatility approximation of \cite{hagan2002managing}, see Appendix \ref{sec:black_hagan} for explicit expressions.\footnote{Remark that \eqref{eq:black_fwd}-\eqref{eq:hagan_fwd} only makes sense for $\tau_0>0$. For $\tau_0 \le 0$, we have $V_{cpl}^f(0)=P(0,\tau_1)(R(\tau_0) -K)^+$.}  Pricing backward-looking caplets, i.e., options on $R(\tau_1)$, is more involved because of the time dependence through the function $\psi(t)$. In the next section we derive effective SABR parameters $\hat{\alpha}$, $\hat{\rho}$, $\hat{\nu}$ such that
\begin{align}
    &V_{cpl}^b(0) = P(0,\tau_1) \pi\left(\tau_1,K,  R(0), \sigma^b_{IV}\right),\label{eq:black_bwd}\\
    &\sigma^b_{IV} \approx \sigma_{hagan}(\tau_1,K,R(0),\hat{\alpha},\beta,\hat{\rho},\hat{\nu}).\label{eq:hagan_bwd}
\end{align}
We leave $\beta$ unchanged, since the local volatility component in the backward-looking SABR model has not changed compared to the standard SABR model. Note the difference between \eqref{eq:black_fwd}-\eqref{eq:hagan_fwd} and \eqref{eq:black_bwd}-\eqref{eq:hagan_bwd} in the time-to-exercise parameter, which reflects the fact that the forward-looking rate fixes at $\tau_0$ while the backward-looking rate fixes at $\tau_1$. Also note that no additional treatment is required for pricing a backward-looking caplet with $\tau_0 < 0$, i.e., when some of the rate fixings in the payoff have already realized. 

We see two practical use cases for the backward-looking SABR model. Suppose first that we observe liquid prices for both backward and forward-looking caplets at a range of strikes. One possibility is to fix $q$ to an exogenous value (e.g., based on historical data), and then use the model to separately mark two sets of parameters $(\alpha, \rho, \nu)$ to backward and forward-looking caplet prices.\footnote{The parameter $\beta$ is typically not calibrated to market prices, but rather to historical data or expert judgement.} The forward-looking parameter markings are just the standard SABR parameters. The backward-looking parameter markings represent the prices of the corresponding forward-looking caplets in the standard SABR model. The purpose of the backward-looking SABR model extension is in this case to make the two sets of marked parameters comparable, since it takes care of the additional volatility in the accrual period under the hood of the model. This allows a trader to easily compare how expensive or cheap the backward-looking caplet is versus the forward-looking one. Alternatively, we can opt to use a single set of parameters for both backward and forward-looking caplets. Suppose for example that prices of forward-looking caplets are more liquid. In this case, we could mark $(\alpha,\rho,\nu)$ to forward-looking caplets, and mark $q$ to hit the at-the-money backward-looking caplet.

\begin{remark}\label{remark:change_ttm}
We choose to derive effective SABR parameters with respect to a time-to-exercise parameter $\tau_1$,  since this is the only choice that allows to price backward-looking caplets throughout the entire accrual period. We note however that it is straightforward to transform these parameters to correspond to a different time-to-exercise parameter $T>0$ using the relationship\footnote{This relationship follows from a simple time-change argument in the SABR model. One can directly verify that it also holds in the implied volatility approximation of \cite{hagan2002managing}.}
\[
\pi(\tau_1,K,R(0),\sigma_{hagan}(\tau_0,\hat{\alpha},\beta,\hat{\rho},\hat{\nu})) = 
\pi\left(T,K,R(0),\sigma_{hagan}\left(T,\hat{\alpha}\sqrt{\frac{\tau_1}{T}},\beta,\hat{\rho},\hat{\nu}\sqrt{\frac{\tau_1}{T}}\right)\right).
\]
For example, when $\tau_0>0$ we can express the effective parameters in terms of a time-to-exercise parameter $\tau_0$ by simply multiplying $\hat{\alpha}$ and $\hat{\nu}$ by $\sqrt{\frac{\tau_1}{\tau_0}}$.
\end{remark}

\begin{remark}
The dynamics in \eqref{eq:spec_R}-\eqref{eq:spec_sig} are in fact a special case of the general FMM model introduced by \cite{lyashenko2019looking} for once specific tenor. Our results can therefore be used to construct a SABR style FMM model with analytic prices for both backward and forward-looking caplets.
\end{remark}

\section{Backward-looking effective SABR parameters}\label{sec:effective}
We apply the results of \cite{hagan2018managing} to find effective SABR parameters $\hat{\alpha}$, $\hat{\rho}$, $\hat{\nu}$ for the model in \eqref{eq:spec_R}-\eqref{eq:spec_sig}. Their results, based on singular perturbation techniques and effective medium theory, are not exact but have the same accuracy as the original SABR approximation of \cite{hagan2002managing}. We distinguish two cases, based on the sign of $\tau_0$. 

If $\tau_0\le 0$, then part of the backward-looking caplet's payoff has realized already. In this case $\psi(t)$ is strictly decreasing towards zero as time moves forward, which decreases the volatility of $R(t)$. The following theorem presents effective SABR parameters incorporating this time-dependent volatility scaling:
\begin{theorem}\label{thm:tau0Negative}
If $\tau_0\le 0$, then the backward-looking effective SABR parameters are
\begin{align*}
&\hat{\rho} = \frac{2\rho}{\sqrt{\zeta}(3q+2)},\quad  \hat{\nu}^2 = \nu^2 \zeta (2q+1),
\quad\hat{\alpha}^2 = \frac{\alpha^2}{2q+1}\left(\frac{\tau_1}{\tau_1-\tau_0}\right)^{2q} \e^{\frac{1}{2}(\frac{\nu^2}{q+1} - \hat{\nu}^2)\tau_1},
\end{align*}
where we have defined
\[
\zeta = \frac{3}{4q+3}\left(\frac{1}{2q+1} + \rho^2\frac{2q}{(3q+2)^2}\right).
\]
\end{theorem}
\begin{proof}
See Appendix.
\end{proof}
Remark that the effective smile parameters $\hat{\rho}$ and $\hat{\nu}$ in Theorem \ref{thm:tau0Negative} do not depend on $\tau_0$ and $\tau_1$, which is linked to a particular scaling property of the backward-looking SABR model when $\tau_0\le 0$. 
%
%
To see this, define the processes $R'(t)$ and $\sigma'(t)$ as
\[
R'(t) = R(\tau_1 t),\quad \sigma'(t)=\sqrt{\tau_1} \left(\frac{\tau_1}{\tau_1-\tau_0}\right)^q\sigma(\tau_1 t).
\]
Using It\^o's lemma and time-scaling properties of the Brownian motion we get for $t\ge \tau_0$
\begin{align*}
&\dd R'(t) = R'(t)^\beta(1-t)^q\sigma'(t)\,\dd B(t),\quad  \dd \sigma'(t) = \nu \sqrt{\tau_1} \sigma'(t)\,\dd W(t).
\end{align*}
Hence, if $\tau_0\le 0$, then the time-0 price of a backward-looking caplet with accrual period $[\tau_0,\tau_1]$ is equivalent to one with canonical accrual period $[0,1]$ together with the following re-parameterization
\begin{equation}
\alpha\to \alpha \sqrt{\tau_1}\left(\frac{\tau_1}{\tau_1-\tau_0}\right)^q,\quad \nu\to\nu\sqrt{\tau_1}.
\label{eq:subs}
\end{equation}
In other words, starting from effective parameters for the accrual period $[0,1]$, one can obtain the effective parameters for any accrual period $[\tau_0,\tau_1]$, $\tau_0\le 0<\tau_1$, by first making the substitutions \eqref{eq:subs} and then adjusting for the change in time-to-exercise from $1$ to $\tau_1$ (see Remark \ref{remark:change_ttm}). One can easily verify that this scaling property is indeed satisfied for the effective parameters in Theorem \ref{thm:tau0Negative}.\footnote{Note that the scaling property is not guaranteed to hold a priori, since the effective parameters in Theorem \ref{thm:tau0Negative} are only approximations. The fact that it does hold is comforting.} In particular, since there is no dependence on $\alpha$ or $\nu$ in $\hat{\rho}$, the effective correlation does not depend on the accrual period. At first sight, the effective volatility-of-volatility may have a dependence on $\tau_1$ through the substitution \eqref{eq:subs}, however this is offset by the subsequent change in time-to-exercise from $1$ to $\tau_1$.

The only effective parameter that depends on how far we are in and how much is remaining of the accrual period is therefore $\hat{\alpha}$. The exponential factor in $\hat{\alpha}$ is a second order correction and does typically not contribute much. At the start of the accrual period, i.e., when $\tau_0=0$, $\hat{\alpha}^2$ is roughly $(2q+1)$ times smaller than $\alpha^2$. The fact that $\hat{\alpha}^2$ is smaller than $\alpha$ is intuitive. Indeed, by increasing the time-to-exercise from $\tau_0$ to $\tau_1$ we increase the option value, ceteris paribus. However, because of the volatility reduction in the accrual period we have increased the option value too much and need to bring it down again with a smaller effective $\hat{\alpha}$. When we price further in the accrual period, i.e., when we decrease $\tau_0$ and $\tau_1$, then $\hat{\alpha}$ gradually decreases towards zero, which reflects the fact that an increasing part of the compounded overnight rate has realized already.


If $\tau_0\ge 0$, then none of the rates referenced in the backward-looking caplet's payoff have been realized yet as of pricing time 0. The rate $R(t)$ follows standard SABR dynamics for $t\le \tau_0$, and only afterwards will the volatility process be scaled down towards zero as $t\to \tau_1$. The following theorem provides the effective SABR parameters in this case.

\begin{theorem}\label{thm:tau0Positive}
If $\tau_0\ge 0$, then the backward-looking effective SABR parameters are
\begin{align*}
&\hat{\rho} = \rho \frac{3\tau^2 + 2q\tau_0^2 + \tau_1^2}{\sqrt{\gamma}(6q+4)},\quad
 \hat{\nu}^2 = \nu^2 \gamma \frac{2q+1}{\tau^3\tau_1},\quad
\hat{\alpha}^2 =\frac{ \alpha^2}{2q+1}\frac{\tau}{\tau_1}\e^{\frac{1}{2}H\tau_1},
\end{align*}
where we have defined
\begin{align*}
& \tau = 2q\tau_0 + \tau_1,\quad H = \nu^2\frac{\tau^2 +2q\tau_0^2 + \tau_1^2}{2\tau_1\tau(q+1)} - \hat{\nu}^2,\\
&\gamma =\tau
\frac{2\tau^3 + \tau_1^3 + (4q^2-2q)\tau_0^3 + 6q\tau_0^2\tau_1}{(4q+3)(2q+1)}+3q\rho^2 (\tau_1-\tau_0)^2\frac{
3\tau^2 - \tau_1^2 +5q\tau_0^2 + 4\tau_0\tau_1}{(4q+3)(3q+2)^2}.
\end{align*}
\end{theorem}
\begin{proof}
See Appendix.
\end{proof}

When pricing at the start of the accrual period, i.e., when $\tau_0 = 0$, the effective parameters in Theorem \ref{thm:tau0Negative} and \ref{thm:tau0Positive} agree. For $\tau_1=\tau_0>0$, the backward-looking SABR model reduces to the standard SABR model and Theorem \ref{thm:tau0Positive} indeed gives in this case $\hat{\rho}=\rho$, $\hat{\nu}=\nu$, and $\hat{\alpha}=\alpha$. 

\begin{figure}[h]
\centering
\begin{subfigure}{.45\textwidth}
  \centering
  \includegraphics[width=\textwidth]{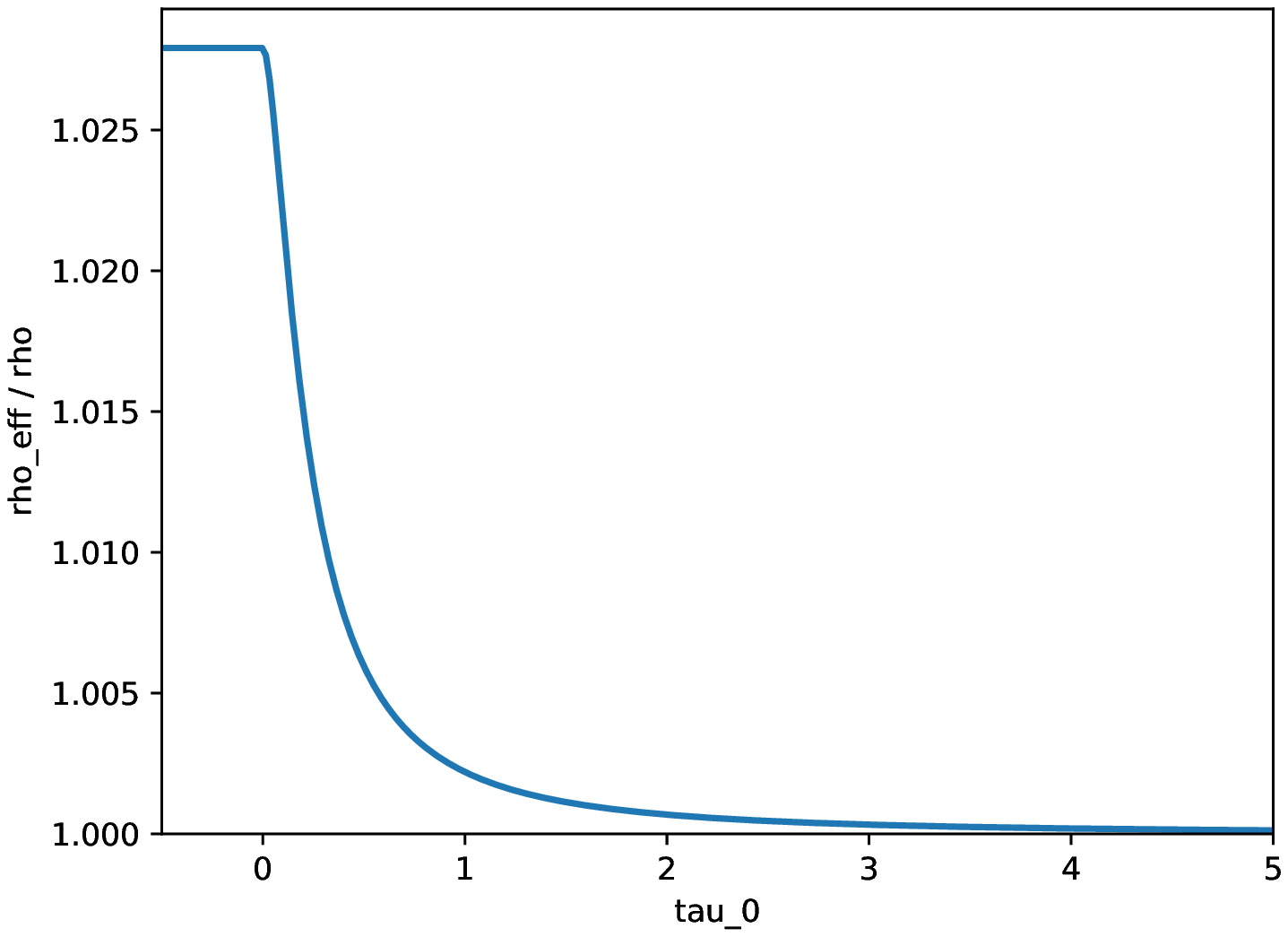}
  \caption{$\hat{\rho}/\rho$}
  \label{fig:rhoEff}
\end{subfigure}
\begin{subfigure}{.45\textwidth}
  \centering
  \includegraphics[width=\textwidth]{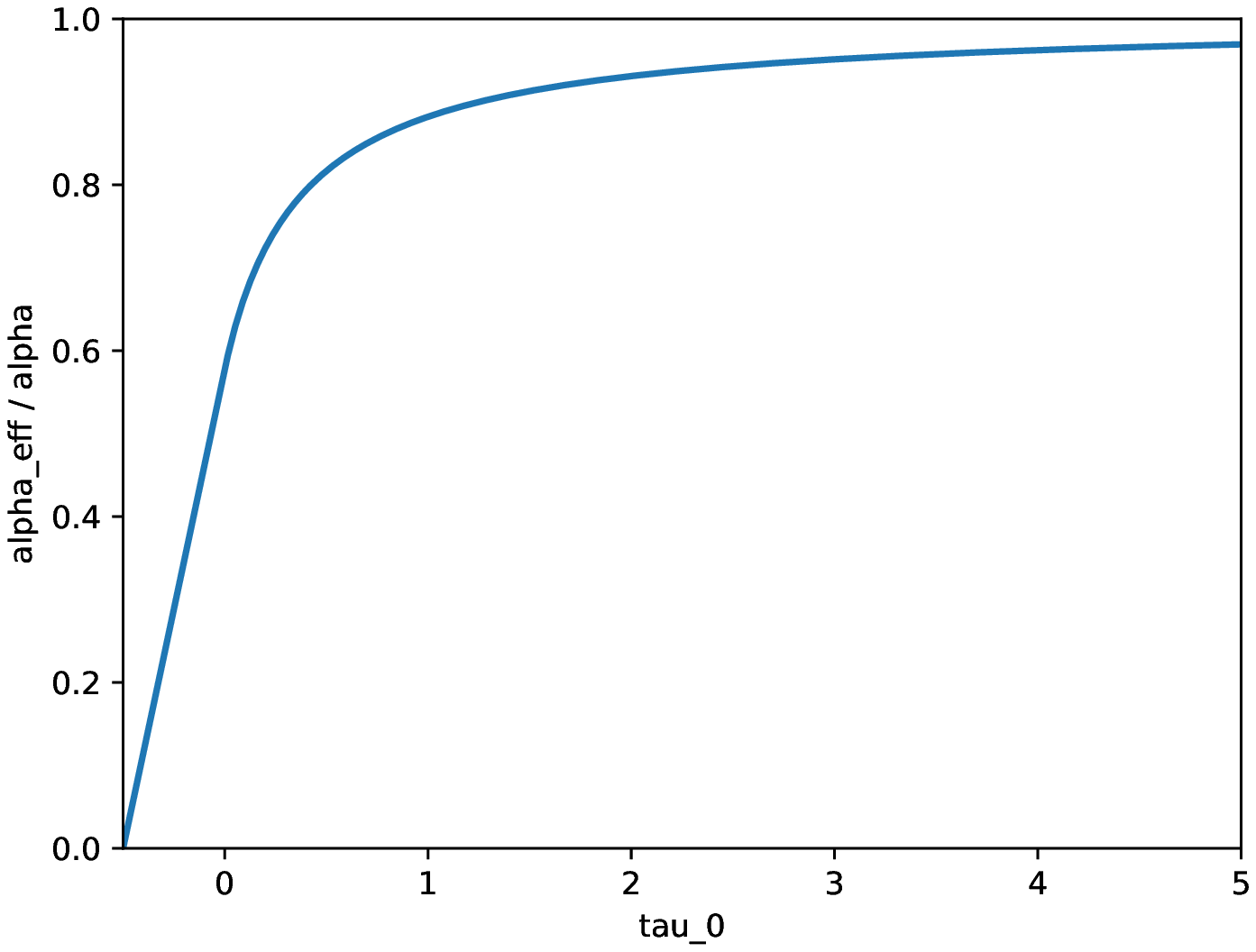}
  \caption{$\hat{\alpha}/\alpha$}
  \label{fig:alphaEff}
\end{subfigure}
\begin{subfigure}{.45\textwidth}
  \centering
  \includegraphics[width=\textwidth]{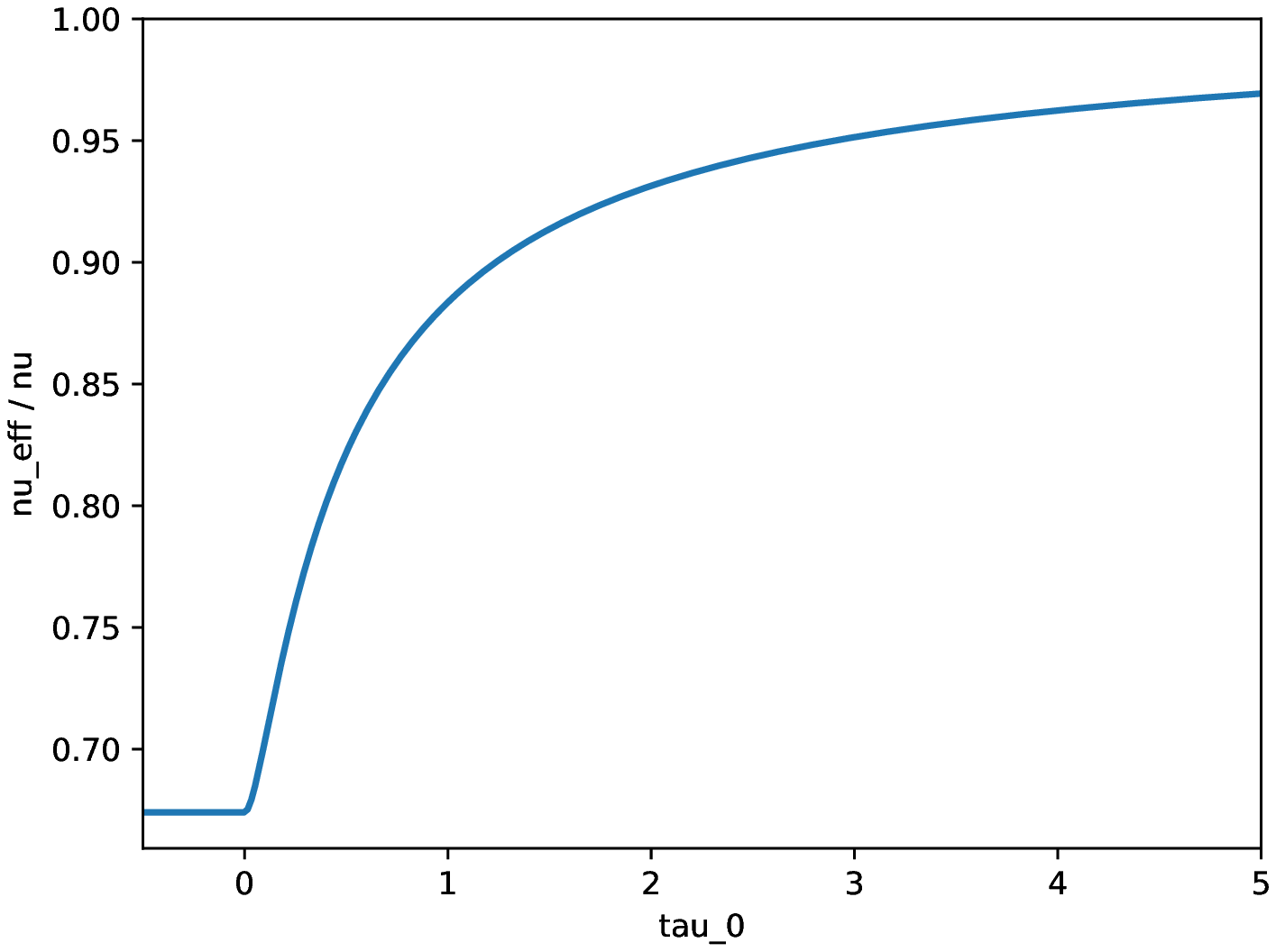}
  \caption{$\hat{\nu}/\nu$}
  \label{fig:nuEff}
\end{subfigure}
\caption{Ratio of effective SABR parameters over original ones for $\tau_0\in[-0.5,5]$. Parameter choices: $\tau_1-\tau_0 = 0.5$, $q=1$, $\rho = -50\%$, $\nu = 50\%$.}
\label{fig:effectiveParams}
\end{figure}

In Figure \ref{fig:effectiveParams}, we set $\tau_1-\tau_0 = 0.5$, $q=1$, $\rho = -50\%$, $\nu = 50\%$, and plot the ratios $\hat{\rho}/\rho$, $\hat{\nu}/\nu$, and $\hat{\alpha}/\alpha$ for $\tau_0\in[-0.5,5]$. We observe that the impact is strongest for small $\tau_0$ and becomes less noticeable for larger $\tau_0$. This makes sense intuitively, since for large $\tau_0$, the model behaves like the standard SABR model most of the time. We also observe that the effective correlation does not change much, while the effective volatility-of-volatility and the initial volatility are reduced substantially for small $\tau_0$.

For $q\to\infty$, the volatility of $R(t)$ is reduced to zero immediately after the start of the accrual period. In this limit case, the backward looking and forward-looking caplet should have the same price. Taking the limit $q\to\infty$ for the effective parameters in Theorem \ref{thm:tau0Positive} gives
\[
\lim_{q\to\infty}\hat{\rho} = \rho,\quad \lim_{q\to\infty}\hat{\nu}^2 = \nu^2 \frac{\tau_0}{\tau_1},\quad \lim_{q\to\infty} \hat{\alpha}^2 = \alpha^2\frac{\tau_0}{\tau_1}. 
\]
Recall that the effective parameters $\hat{\alpha},\hat{\rho},\hat{\nu}$ correspond to a time-to-exercise parameter $\tau_1$, while $\alpha, \rho, \nu$ correspond to a time-to-exercise parameter $\tau_0$. From the observation in Remark \ref{remark:change_ttm} we immediately see that the backward and forward-looking caplet indeed have equal price in this limit case. Taking the limit $q\to\infty$ for the effective parameters in Theorem \ref{thm:tau0Negative} gives zero values for all effective parameters, which is not surprising since there is no volatility left in the accrual period. For $q\to 0$, the volatility decays very slow in the accrual period. Taking the limit $q\to 0$ for the effective parameters in Theorem \ref{thm:tau0Negative} and Theorem \ref{thm:tau0Positive} gives
\[
\lim_{q\to 0}\hat{\rho} = \rho,\quad \lim_{q\to 0}\hat{\nu}^2 = \nu^2,\quad \lim_{q\to 0} \hat{\alpha}^2 = \alpha^2.
\]
The effective parameters therefore coincide with the original ones. However, we note again that because of the difference in the corresponding time-to-exercise parameter, this does not mean that the backward and forward-looking caplet have the same price. In fact, in the limit case $q\to 0$, the difference between the two is maximized.

\begin{remark}
\cite{piterbarg2020interest} derives for $\tau_0>0$ an adjusted initial volatility 
\begin{equation}
    \tilde{\alpha} = \alpha\sqrt{1+\frac{\tau_1 -\tau_0}{3\tau_0}}\label{eq:piterbarg_alpha}
\end{equation}
such that
\begin{align*}
    &V_{cpl}^b(0) = P(0,\tau_1)\pi(\tau_0,K,R(0),\tilde{\sigma}_{IV}^b),\\
    &\tilde{\sigma}_{IV}^b \approx \sigma_{hagan}(\tau_0, K, R(0), \tilde{\alpha}, \beta, \rho, \nu).
\end{align*}
Note that $\tilde{\alpha}$ corresponds to a time-to-exercise $\tau_0$. In order to compare $\hat{\alpha}$ with $\tilde{\alpha}$, we need to scale it appropriately (see Remark \ref{remark:change_ttm}):
\begin{align}
    \hat{\alpha}\sqrt{\frac{\tau_1}{\tau_0}} = \alpha\sqrt{\frac{\tau}{(2q+1)\tau_0}} \e^{\frac{1}{4}H\tau_1}
    = \alpha\sqrt{1+\frac{\tau_1 -\tau_0}{(2q+1)\tau_0}}\e^{\frac{1}{4}H\tau_1}.\label{eq:piterbarg_adjustment}
\end{align}
Ignoring the exponential factor, which is typically very close to one, and setting $q=1$, we observe that \eqref{eq:piterbarg_alpha} agrees with \eqref{eq:piterbarg_adjustment}. 
\end{remark}

\section{Numerical study}\label{sec:numerical}
\begin{figure}
    \centering
    \includegraphics[width=0.8\textwidth]{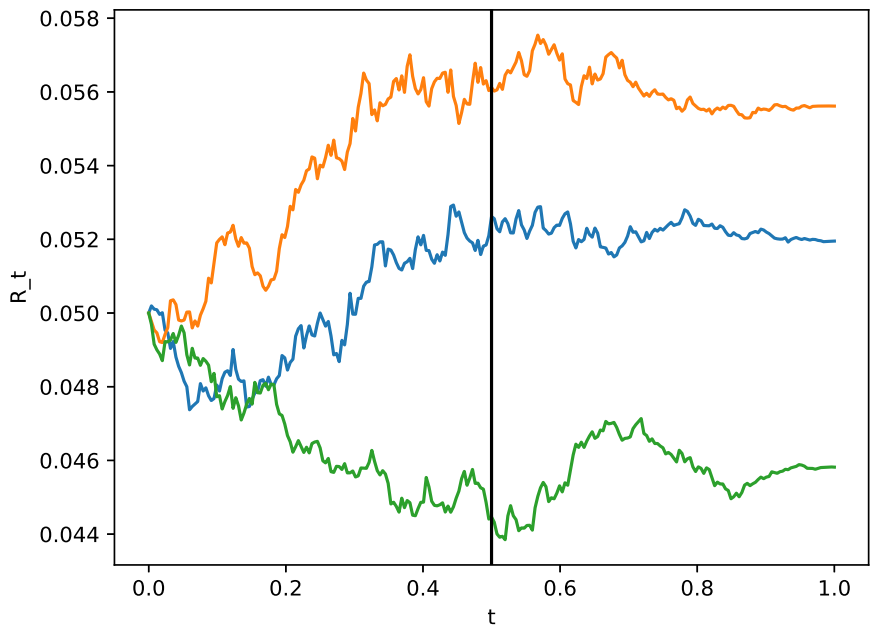}
    \caption{Three simulated trajectories of $R(t)$. The vertical black line indicates the start of the accrual period $t=\tau_0$.}
    \label{fig:simulated_rate}
\end{figure}

\begin{figure}
    \centering
    \includegraphics[width=0.8\textwidth]{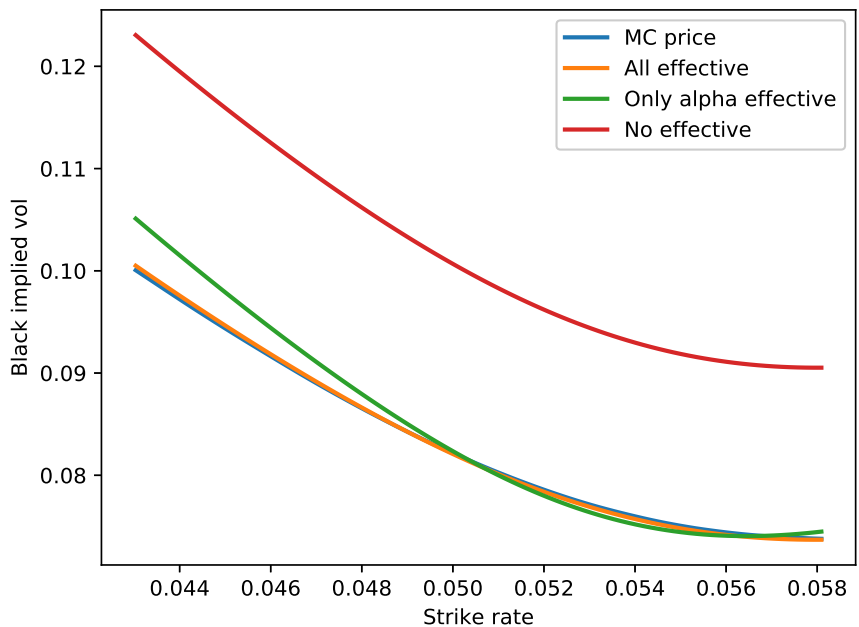}
    \caption{Black implied volatility smiles. The blue line corresponds to the Monte-Carlo price, the orange line to the SABR price with $(\hat{\alpha}, \hat{\rho}, \hat{\nu})$, the green line to SABR price with $(\hat{\alpha}, \rho, \nu)$, and the red line to the SABR price with $(\alpha, \rho, \nu)$.}
    \label{fig:MC_smile}
\end{figure}

In order to evaluate the approximation quality of the effective SABR parameters derived in the previous section, we perform a Monte-Carlo simulation of the model as a benchmark. We arbitrarily choose the following parameter values: $\beta=1$, $R(0) = 0.05$, $\tau_0 = 0.5$, $\tau_1 = 1$, $\rho = -0.50$, $\nu = 0.50$, $\alpha = 0.10$, and $q=1$. The effective parameters, computed using Theorem \ref{thm:tau0Positive}, become
\[
\hat{\alpha} = 0.082,\quad \hat{\rho} = -0.503,\quad \hat{\nu} = 0.411.
\]
We simulate \eqref{eq:spec_R}-\eqref{eq:spec_sig} using a log-Euler discretization scheme with time step $1/512$ and $10^6$ simulation trajectories. Figure \ref{fig:simulated_rate} shows an example of three simulated trajectories of $R(t)$. Notice how the volatility starts to decrease for $t>\tau_0$. We use the simulated trajectories to price backward-looking caplets and then compute Black implied volatility by inverting \cite{black1976pricing}'s formula. Figure \ref{fig:MC_smile} shows that \cite{hagan2002managing}'s implied volatility approximation with effective parameters
$\hat{\alpha}, \hat{\rho}, \hat{\nu}$ produces virtually identical results as the Monte-Carlo simulation. As a reference, we also plot implied volatilities computed using  \cite{hagan2002managing}'s approximation with parameters $\alpha, \rho, \nu$ and $\hat{\alpha}, \rho, \nu$. In the first case, we do not apply any adjustments to the SABR parameters, and Figure \ref{fig:MC_smile} clearly shows we are substantially mispricing caplets at all strikes. In the second case, we only apply a correction to the initial volatility parameter. The at-the-money strike is priced correctly in this case, but strikes away from the money are not. 

\section{Conclusion}\label{sec:conclusion}
We have presented a natural extension of the SABR model to consistently price backward and forward-looking caplets. Forward-looking caplets can be priced using the market standard approximations of \cite{hagan2002managing}. Building on the results of \cite{hagan2018managing}, we have derived closed-form effective SABR parameters for pricing backward-looking caplets. We have shown the effective SABR approximation to be highly accurate when compared to a Monte-Carlo simulation benchmark. The correction to the initial volatility and the volatility-of-volatility parameters are most pronounced, while the correction to the correlation parameter is of second order importance. Our results allow traders to think of backward and forward-looking caplets in normalized units, which helps to assess their relative values. A similar analysis can be done for the \cite{heston1993closed} model using, for example, the results from \cite{hagan2018implied}. We leave such extensions for future research.
\clearpage

\appendix

\section{Term rate implied by Hull-White model}\label{app:short-rate}
In this section we investigate what dynamics are implied for $R(t)$ from the \cite{hull1990pricing} short-rate model. The goal of this exercise is to gain insight in what happens to the volatility of $R(t)$ inside the accrual period. Specifically, consider the following risk-neutral dynamics for the short-rate $r(t)$:
\begin{align*}
    \dd r(t)& = (\theta(t) - \kappa r(t))\,\dd t + \xi(t)\, \dd B^\Q(t),
\end{align*}
where $\theta(t)$ is a deterministic function chosen such that the initial term-structure $T\mapsto P(0,T)$ is recovered, $\xi(t)$ is a deterministic volatility function, $B^\Q(t)$ is a standard Brownian motion under the risk-neutral measure $\Q$ where the money-market account serves as numeraire, and $\kappa\in\R$ is the speed of mean-reversion. The time-$t$ price of a zero-coupon bond with maturity $T\ge t$ is given by
\begin{equation*}
    P(t,T) = \e^{-C(t,T) - D(t,T)r(t)},\quad D(t,T) = \frac{1 - \e^{-\kappa (T - t)}}{\kappa},
\end{equation*}
and $C(t,T)$ is a deterministic function that plays no particular role here. Taking the short-rate as an instantaneous proxy of the overnight rate, we get
\begin{align*}
1+(\tau_1 - \tau_0) R(t) &=
\frac{1}{P(t,\tau_1)}\E_t^\Q\left[\e^{-\int_t^{\tau_1}r(s)\,\dd s}\e^{\int_{\tau_0}^{\tau_1}r(s)\,\dd s}\right]\\
&=
\frac{1}{P(t,\tau_1)} \times 
\begin{cases}
P(t,\tau_0), & t\le \tau_0,\\
\e^{\int_{\tau_0}^{t}r(s)\,\dd s}, & t\ge \tau_0,
\end{cases}
\end{align*}
where $\E_t^\Q[\cdot]$ denotes the conditional expectation under $\Q$. 
Standard arguments show that the dynamics of $R(t)$ under the $\tau_1$-forward measure become
\begin{equation}
\dd R(t)= \left(\frac{1}{\tau_1 - \tau_0} + R(t)\right) \times
\begin{cases}
(D(t,\tau_1) - D(t,\tau_0))\,\dd B(t), & t\le \tau_0,\\
D(t,\tau_1)\,\dd B(t), & t\ge \tau_0,
\end{cases}
\label{eq:HW_R}
\end{equation}
where $B(t)$ is a standard Brownian motion under the $\tau_1$-forward measure.
Defining
\begin{align*}
   & \tilde{\psi}(t) =  \min\left(1,\frac{\e^{-\kappa t} - \e^{-\kappa \tau_1}}{\e^{-\kappa \tau_0} - \e^{-\kappa \tau_1}}\right),\quad 
   \tilde{\sigma}(t) = \frac{\e^{-\kappa(\tau_0-t)} - \e^{-\kappa(\tau_1-t)}}{\kappa}\xi(t),
\end{align*}
we can rewrite \eqref{eq:HW_R} as
\begin{equation*}
    \dd R(t) = \left(\frac{1}{\tau_1 - \tau_0} + R(t)\right) \tilde{\psi}(t) \tilde{\sigma}(t)\, \dd B(t).
\end{equation*}
The function $\tilde{\psi}(t)$ is equal to one for $t\in[0,\tau_0]$ and decays monotonically towards zero for $t\in[\tau_0,\tau_1]$. The speed of mean-reversion parameter $\kappa$ controls how fast $\tilde{\psi}(t)$ decreases. The qualitative behaviour of the function $\psi(t)$ defined in \eqref{eq:spec_R} is very similar to $\tilde{\psi}(t)$, where $q-1$ plays the role of $\kappa$. Indeed, $\psi(t)$ and $\tilde{\psi}(t)$ are both equal to one for $t\in[0,\tau_0]$ and decay monotonically towards zero for $t\in[\tau_0,\tau_1]$. For $\kappa>0$ and $q>1$, the decay is convex, for $\kappa<0$ and $0<q<1$ it is concave, and for $\kappa=0$ and $q=1$ it is linear. The reason we did not use $\tilde{\psi}(t)$ as a definition in the backward-looking SABR model is that the expressions for the effective SABR parameters become much more complicated. However, this example shows that $q$ plays a similar role to the speed of mean-reversion in a short-rate model.

\section{\cite{black1976pricing} and \cite{hagan2002managing} formulas}\label{sec:black_hagan}
In the main text we have used \cite{black1976pricing}'s formula and the Black implied volatility approximation for the SABR model by \cite{hagan2002managing}. Both are well known and standard tools in interest rate derivatives markets, however for the sake of completeness we write them out explicitly in this section. 

The \cite{black1976pricing} formula is 
\[
\pi(T,K,R,\sigma) = R\Phi(d_+) - K\Phi(d_-),\quad d_\pm = \frac{x \pm \frac{\sigma^2}{2}T}{\sigma \sqrt{T}},
\]
with $x=\log\left(\frac{R}{K}\right)$, $\Phi(\cdot)$ the standard normal cumulative distribution, and $R,K,T,\sigma>0$.  

The Black implied volatility approximation for the SABR model by \cite{hagan2002managing} is
\begin{align*}
    \sigma_{hagan}(T,K,R,\alpha,\beta,\rho,\nu)=&
    \frac{\alpha}{(RK)^{(1-\beta)/2}}\left(1+\frac{(1-\beta)^2}{24}x^2 + \frac{(1-\beta)^4}{1920}\right)^{-1}
    \frac{z}{\chi(z)}\\
    &\left(1+\left[\frac{(1-\beta)^2}{24}\frac{\alpha^2}{(RK)^{1-\beta}} + \frac{1}{4}\frac{\rho\beta\nu\alpha}{(RK)^{(1-\beta)/2}} + \frac{2-3\rho^2}{24}\nu^2\right]T\right),
\end{align*}
with 
\[
z = \frac{\nu}{\alpha} (RK)^{(1-\beta)/2}x,\quad \chi(z) = \log\left(\frac{\sqrt{1-2\rho z+ z^2} + z - \rho}{1-\rho}\right).
\]

\section{Proofs}
\subsection{Proof of Theorem \ref{thm:tau0Negative}}
We start by summarizing a particular case of the results in \cite{hagan2018managing}.
\begin{theorem}[\cite{hagan2018managing}]\label{thm:hagan}
Consider for $\epsilon>0$ the following dynamics
\begin{align}
    &\dd R_\epsilon (t) = \epsilon\phi(t)R_\epsilon(t)^\beta \sigma_\epsilon(t) \,\dd B(t) \label{eq:dyn_SABR_1} \\
    &\dd \sigma_\epsilon (t) = \epsilon\nu \sigma_{\epsilon}(t)\, \dd W(t), \quad \sigma_\epsilon(0) = 1. \label{eq:dyn_SABR_2}
\end{align}
For a given expiry $T > 0$, \cite{hagan2018managing} show that to within $\Ocal (\epsilon^2)$, the implied volatility of European options under the model in \eqref{eq:dyn_SABR_1}-\eqref{eq:dyn_SABR_2} coincides with the implied volatility under the standard SABR model with constant parameters
\[
\hat{\alpha} = \Delta \e^{\frac{1}{4}\epsilon^2 \Delta^2 G T},\quad \hat{\rho} = \frac{b}{\sqrt{c}},\quad \hat{\nu} = \Delta \sqrt{c},
\]
where the following constants are defined
\begin{align*}
    &\Delta^2 = \frac{v(T)}{T}, \quad 
    b = \frac{2 \rho \nu}{v^2(T)} \int_0^T (v(T) - v(s))\phi(s)\, \dd s,\quad
    G = \frac{2\nu^2}{v^2(T)}\int_0^T v(T) - v(s) \, \dd s - c,\\
    &c = \frac{3 \nu^2}{v^3(T)}\int_0^T(v(T) - v(s))^2\, \dd s + \frac{9}{v^3(T)}\int_0^T w^2(s)\phi^2(s)\,\dd s - 3b^2,
\end{align*}
with $v\colon u\mapsto \int_0^u \phi^2(s)\,\dd s$, and $w\colon u\mapsto \rho\nu\int_0^u \phi(s)\,\dd s$.
\end{theorem}
Note that the accuracy of the original SABR approximation in \cite{hagan2002managing} was also of the order $\Ocal(\epsilon^2)$. The accuracy of the effective parameter approximation is therefore of the same order as the original approximation.

We can apply the above result to the backward-looking SABR model in \eqref{eq:spec_R}-\eqref{eq:spec_sig} by setting $\epsilon = 1$, $\phi(t) = \alpha \psi(t)$, and $T=\tau_1$. For $\tau_0<0$, we have therefore
\[
\phi(t) = \alpha \left(\frac{\tau_1 - t}{\tau_1 - \tau_0}\right)^q,\quad t\le \tau_1.
\]
It remains to calculate the integrals appearing in Theorem \ref{thm:hagan}. We start with the functions $v$ and $w$:
\begin{align*}
    &v(u) = \frac{\alpha^2}{(\tau_1-\tau_0)^q}\int_0^u (\tau_1-s)^{2q}\,\dd s = \alpha^2\frac{\tau_1^{2q+1} - (\tau_1-u)^{2q+1}}{(2q+1)(\tau_1-\tau_0)^{2q}},\\
    & w(u) = \frac{\rho\nu}{\tau_1-\tau_0} \int_0^u (\tau_1 - s)\,\dd s =\rho\nu\alpha \frac{\tau_1^{q+1} - (\tau_1 - u)^{q+1}}{(q+1)(\tau_1-\tau_0)^q}.
\end{align*}
In particular, we have $v(\tau_1) =\alpha^2 \frac{\tau_1^{2q+1}}{(2q+1)(\tau_1-\tau_0)^{2q}}$ and $v(\tau_1)-v(s) = \alpha^2 \frac{(\tau_1 - s)^{2q+1}}{(2q+1)(\tau_1-\tau_0)^{2q}}$. Elementary, but rather tedious, calculations give
\begin{align*}
    &b = \frac{\nu\rho(4q+2)}{\alpha(3q+2)}\left(\frac{\tau_1-\tau_0}{\tau_1}\right)^q,\\
    & c = \frac{\nu^2}{\alpha^2}\left(\frac{\tau_1-\tau_0}{\tau_1}\right)^{2q}\frac{2q+1}{(3q+2)^2(4q+3)}\left((3q+2)^2+ \rho^2(4q^2+2q)\right),\\
    &\Delta^2 = \frac{\alpha^2}{2q+1} \left(\frac{\tau_1}{\tau_1 - \tau_0}\right)^{2q},\quad 
    G  = \frac{\nu^2}{\Delta^2(q+1)} - c.
\end{align*}
Putting things together gives
\begin{align*}
    &\hat{\nu}^2 = \Delta^2 c =  \nu^2 \zeta (2q+1),\\
    &\hat{\rho} = \frac{b}{\sqrt{c}} = \frac{2\rho}{\sqrt{\zeta}(3q+2)},\\
    &\hat{\alpha}^2 =\Delta^2 \e^{\frac{1}{2} \Delta^2 G \tau_1} = \frac{\alpha^2}{2q+1}\left(\frac{\tau_1}{\tau_1-\tau_0}\right)^{2q} \e^{\frac{1}{2}(\frac{\nu^2}{q+1} - \hat{\nu}^2)\tau_1},
\end{align*}
with $\zeta$ as defined in Theorem \ref{thm:tau0Negative}.

\subsection{Proof of Theorem \ref{thm:tau0Positive}}
Similarly as in the proof of Theorem \ref{thm:tau0Negative}, we again apply Theorem \ref{thm:hagan} with $\epsilon=1$, $\phi(t) = \alpha \psi(t)$, and $T=\tau_1$. For the case $\tau_0\ge 0$ we get
\[
\phi(t) = 
\begin{cases}
\alpha, & 0\le t\le \tau_0,\\
\alpha \left(\frac{\tau_1 - t}{\tau_1 - \tau_0}\right)^q, & \tau_0\le t\le \tau_1.
\end{cases}
\]
The functions $v$ and $w$ become:
\[
v(u) =
\begin{cases}
\alpha^2 u, & 0\le u\le \tau_0,\\
\frac{\alpha^2}{2q +1} \left(2q\tau_0 + \tau_1-\frac{(\tau_1 - u)^{2q+1}}{(\tau_1 - \tau_0)^{2q}}\right), & \tau_0\le u\le \tau_1,
\end{cases}
\]
\[
w(u) =
\begin{cases}
\rho\nu\alpha u, & 0\le u\le \tau_0,\\
\frac{\rho\nu\alpha}{q+1} \left(q\tau_0 + \tau_1-\frac{(\tau_1 - u)^{q+1}}{(\tau_1 - \tau_0)^q}\right), & \tau_0\le u\le \tau_1.
\end{cases}
\]
Elementary, but rather tedious, calculations give
\begin{align*}
   &b = \frac{\nu \rho}{\alpha}(2q+1) \frac{3\tau^2 +2q\tau_0^2 +\tau_1^2}{2\tau^2(3q+2)},\quad
   c = \frac{\nu^2}{\alpha^2}\frac{\gamma (2q+1)^2}{\tau^4}\\
   &\Delta^2 = \frac{\alpha^2\tau}{\tau_1(2q+1)},\quad  
   G = \frac{\nu^2}{\Delta^2}\frac{\tau^2 +2q \tau_0^2 +\tau_1^2}{2\tau_1\tau (q+1)}- c,
\end{align*}
with $\tau$ and $\gamma$ as defined in Theorem \ref{thm:tau0Positive}.

Putting things together gives
\begin{align*}
    &\hat{\alpha}^2 =\Delta^2 \e^{\frac{1}{2}\Delta^2 G \tau_1} =  \frac{\alpha^2\tau}{\tau_1(2q+1)} \e^{\frac{1}{2}\Delta^2 G \tau_1},\\
    &\hat{\nu}^2 = \Delta^2 c = \frac{\nu^2\gamma (2q+1)}{\tau^3\tau_1},\\
    &\hat{\rho} = \frac{b}{\sqrt{c}} = \frac{\rho}{\sqrt{\gamma}} \frac{3\tau^2 +2q\tau_0^2 +\tau_1^2}{2(3q+2)}. 
\end{align*}
Substituting $H=\Delta^2 G$ concludes the proof.
\clearpage
\bibliography{references}
\bibliographystyle{chicago}
\end{document}